\documentclass[showkeys,aps,amsmath,amssymb,twocolumn,groupedaddress,longbibliography,floats,final,postprint,superscriptaddress]{revtex4-1}
\usepackage{enumerate}
\usepackage{graphicx}
\usepackage{amsmath,amssymb,amsthm,enumitem}

\usepackage{color}
\definecolor{LinkColor}{RGB}{199,21,133}
\usepackage[colorlinks=true,linkcolor=blue,anchorcolor=blue,citecolor=blue,urlcolor=blue]{hyperref}
\usepackage[polutonikogreek,english]{babel}

\usepackage{hyperref}

\usepackage{epstopdf}
\usepackage{type1cm}
\usepackage{multirow}
\usepackage{times}

\def\qq{q}

\def\scrh{\mathcal{H}}

\def\scro{\mathcal{O}}

\begin{document}
\author{Jian-Ping Lv}
\email{jplv2014@ahnu.edu.cn}
\author{Wanwan Xu}
\author{Yanan Sun}
\affiliation{Department of Physics, Anhui Key Laboratory of Optoelectric Materials Science and Technology, Key Laboratory of Functional Molecular Solids, Ministry of Education, Anhui Normal University, Wuhu, Anhui 241000, China}
\author{Kun Chen}
\email{chenkun0228@gmail.com}
\affiliation{Department of Physics and Astronomy, Rutgers, The State University of New Jersey, Piscataway, New Jersey 08854-8019}
\author{Youjin Deng}
\email{yjdeng@ustc.edu.cn}
\affiliation{National Laboratory for Physical Sciences at Microscale and Department of Modern Physics, University of Science and Technology of China, Hefei, Anhui 230026, China}
\affiliation{Department of Physics and Electronic Information Engineering, Minjiang University, Fuzhou, Fujian 350108, China}

\title{Finite-size Scaling of O($n$) Systems at the Upper Critical Dimensionality}
\thanks{Author Contributions: J.-P.L., K.C., and Y.D. designed the research and established the formulae for finite-size scaling. J.-P.L., W.X., and Y.S. performed simulations. J.-P.L., W.X., Y.S., and Y. D. analyzed the results. J.-P.L. and Y.D. wrote the manuscript. All the authors participated in the revisions of the manuscript.}
\date{\today}

\begin{abstract}
Logarithmic finite-size scaling of the O($n$) universality class at the upper critical dimensionality ($d_c=4$) has
a fundamental role in statistical and condensed-matter physics and important applications in various experimental
systems. Here, we address this long-standing problem in the context of the $n$-vector model ($n=1, 2, 3$)
on periodic four-dimensional hypercubic lattices.
We establish an explicit scaling form for the free energy density, which simultaneously consists of a scaling term
for the Gaussian fixed point and another term with multiplicative logarithmic corrections.
In particular, we conjecture that the critical two-point correlation $g(r,L)$, with $L$ the linear size, exhibits a two-length behavior:
following the behavior $r^{2-d_c}$ governed by Gaussian fixed point at shorter distance and entering a plateau at larger distance whose height decays as $L^{-d_c/2}({\rm ln}L)^{\hat{p}}$ with $\hat{p}=1/2$ a logarithmic correction exponent.
Using extensive Monte Carlo simulations, we provide complementary evidence for the predictions through the finite-size scaling of observables including the two-point correlation, the magnetic fluctuations at zero and non-zero Fourier modes, and the Binder cumulant. Our work sheds light on the formulation of logarithmic finite-size scaling and has practical applications in experimental systems.
\end{abstract}


\maketitle

\section{Introduction}
The O($n$) model of interacting vector spins is a much-applied model in
condensed-matter physics and one of the most significant classes of lattice models in equilibrium statistical mechanics~\cite{fernandez2013random,svistunov2015superfluid}. The Hamiltonian of the O($n$) vector model is written as
\begin{equation}\label{HamOn}
\scrh=-\sum\limits_{\langle {\bf r}{\bf r'}\rangle}\vec{S}_{\bf r} \cdot \vec{S}_{\bf r'},
\end{equation}
where $\vec{S}_{\bf r}$ is an $n$-component isotropic spin with unit length
and the summation runs over nearest neighbors. Prominent examples include the Ising ($n \! = \! 1$), XY ($n \! = \! 2$) and Heisenberg ($n \! = \! 3$) models of ferromagnetism, as well as the self-avoiding random walk ($n \! \rightarrow \! 0$) in polymer physics.
Its experimental realization is now available for various $n$ values in magnetic materials~\cite{merchant2014quantum,Qin2015,lohofer2017excitation,qin2017amplitude,Cui2019}, superconducting arrays~\cite{Resnick1981,goldman2013percolation} and ultracold atomic systems~\cite{Greiner,Capogrosso-Sansone}.

Finite-size scaling (FSS) is an extensively utilized method for studying systems of continuous phase transitions~\cite{cardy2012finite},
including the O$(n)$ vector model~(\ref{HamOn}).
Near criticality, these systems are characterized by a diverging correlation length $\xi \propto t^{-\nu}$,
where parameter $t$ measures the deviation from the critical point and $\nu$ is a critical exponent.
For a finite box with linear size $L$, the standard FSS hypothesis assumes
that $\xi $ is bounded by the linear size $L$,
and thus predicts that the singular part $f(t,h)$ of free energy density scales as
\begin{equation}\label{eq:f_at_dc24}
\begin{split}
f(t,h) = & L^{-d} \tilde{f} (t L^{y_t}, hL^{y_h})
\end{split}
\end{equation}
where $\tilde{f}$ is a universal scaling function,
$t$ and $h$ represent  the thermal and magnetic scaling fields, and $y_t=1/\nu$ and $y_h$
are the corresponding thermal and magnetic renormalization exponents, respectively.
%the renormalization critical exponents  and $y_h$ correspond to the relevant fields $t$ and $h$, respectively.
Further, the standard FSS theory hypothesizes that at criticality, the spin-spin
correlation function $g(r,L) \equiv \langle \vec{S}_0 \cdot \vec{S}_{\bf r} \rangle $ of
distance $r$ decays as
\begin{equation}\label{g1}
g(r,L) \asymp  r^{-(d-2+\eta)}\tilde{g} (r/L) \; ,
\end{equation}
where $\eta$ relates to $y_h$ by the scaling relation $\eta \! = \! 2+d-2y_h$.
From (\ref{eq:f_at_dc24}) and (\ref{g1}), the FSS of various macroscopic physical quantities can be obtained.
For instance, from the second derivative of $f(t,h)$ with respect to $t$ or $h$,
it is derived that at criticality, the specific heat behaves as  $C \! \asymp \! L^{2y_t-d} $
and the magnetic susceptibility diverges as $\chi \! \asymp \! L^{2y_h-d} $.
The FSS of $\chi$ can also be calculated by summing $g(r,L)$ over the system.
Further, the thermodynamic critical exponents can be obtained
by the (hyper-)scaling relations. For instance, in the thermodynamic limit ($L \rightarrow \infty$),
the specific heat and the magnetic susceptibility scale as $C \propto t^{-\alpha}$
and $\chi \propto t^{-\gamma}$, where the critical exponents are $\alpha = 2-d/y_t$ and $\gamma \! = \! (2y_h-d)/y_t$.

The O($n$) model exhibits an upper critical dimensionality $d_c \! = \! 4$ such that
the thermodynamic scaling in higher dimensions $d \! > \! d_c$ are governed by the Gaussian fixed point,
which has the critical exponents $\alpha=0$ and $\gamma=1$ etc.
In the framework of renormalization group,
the renormalization exponents near the Gaussian fixed point are
\begin{equation} \label{eq:gaussian}
y_t  = 2 \hspace{5mm} \mbox{and} \hspace{5mm}  y_h  = 1 +  d/2
\end{equation}
for $d>d_c$.

Accordingly, the standard FSS formulae~(\ref{eq:f_at_dc24}) and (\ref{g1}) predict that
the critical susceptibility diverges as $\chi \! \asymp \! L^{2y_h-d} \! = \! L^2$ for $d > d_c$.
However, for the Ising model on 5d periodic hypercubes, $\chi$ was numerically observed to scale as $L \asymp L^{5/2}$ instead of $L^2$~\cite{luijten1997interaction,luijten1999finite,Wittmann2014,kenna2014fisher,Flores-Sola,grimm2017geometric}.
The FSS for $d \geq d_c$ turns out to be surprisingly subtle and remains a topic of extensive controversy~\cite{luijten1997interaction,luijten1999finite,papathanakos2006finite,Wittmann2014,kenna2014fisher,Flores-Sola,grimm2017geometric,zhou2018random,fang2019}.

\begin{figure*}
 \includegraphics[width=18cm,height=5cm]{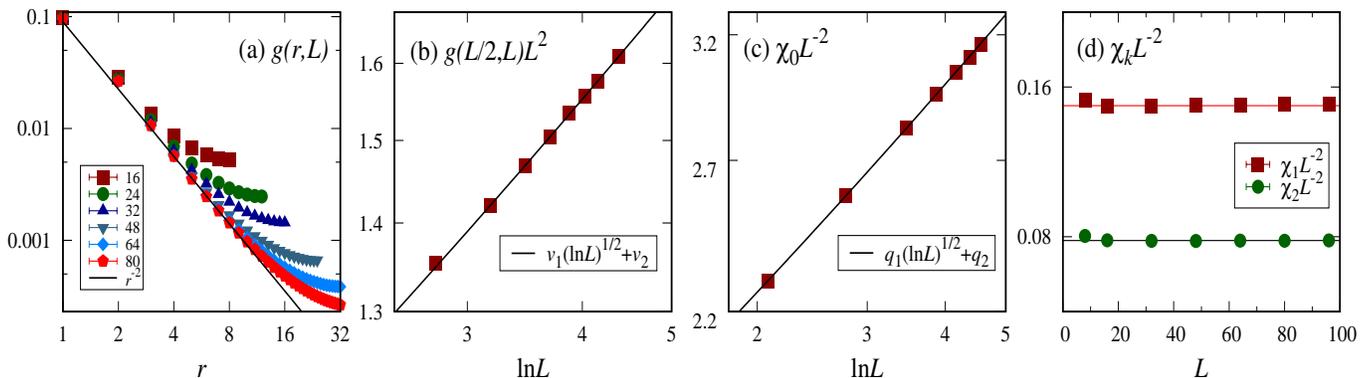}
\caption{Evidence for conjectured formulae  (\ref{eq:f_at_dc}) and (\ref{tp}) in the example of the critical 4d XY model.
(a) correlation function $g(r,L)$ in a log-log scale.
The solid line stands for the $r^{-2}$ behavior.
(b) scaled correlation $g(r,L)L^{2}$ with $r \! = \! L/2$ versus ${\rm ln}L$ in a log-log scale. Thus, the horizontal axis is effectively in a double logarithmic scale of $L$. The solid line represents logarithmic divergence with $\hat{p} \! = \! 1/2$.
(c) scaled magnetic susceptibility $\chi_0 L^{-2}$  versus ${\rm ln}L$ in a log-log scale.
The solid line accounts for logarithmic divergence with $\hat{p} \! = \! 1/2$.
(d) scaled ${\mathbf k} \neq 0$ magnetic fluctuations $\chi_1 L^{-2}$ and $\chi_2 L^{-2}$,
 with ${\mathbf k}_1 \! = \!  (2\pi/L, 0, 0, 0)$ and ${\mathbf k}_2 \! = \! (2\pi/L, 2\pi/L, 0, 0)$, respectively.
 The horizontal lines strongly indicate the absence of logarithmic corrections in the scaling of $\chi_{\bf k}$.}~\label{fig1}
 \end{figure*}

It was realized that for $d > d_c$, the Gaussian exponents $y_t$ and $y_h$ in~(\ref{eq:gaussian})
can be renormalized by the leading irrelevant thermal field with exponent $y_u=4-d$ as~\cite{fisher1971critical,binder1985finite,berche2012hyperscaling,kenna2013new}
%More precisely speaking, the scenario of dangerously irrelevant field (DIF) predicts
\begin{equation}\label{ystar}
y_t^{*}=y_t-\frac{y_u}{2} = \frac{d}{2}  \hspace{2mm} \mbox{and} \hspace{2mm}   y_h^{*}=y_h-\frac{y_u}{4} =\frac{3d}{4} \; ,
\end{equation}
 and the FSS of the free energy density $f(t,h)$ becomes
\begin{equation}\label{eq:f_at_dc23}
\begin{split}
f(t,h) = & L^{-d} \tilde{f} (t L^{y_t^{*}}, hL^{y_h^{*}}).
\end{split}
\end{equation}
In this scenario of dangerously irrelevant field, the FSS of the critical susceptibility becomes $\chi \! \asymp \! L^{2y_h^*-d} \! = \! L^{d/2}$,
consistent with the numerical results~\cite{luijten1997interaction,luijten1999finite,Wittmann2014,Flores-Sola,grimm2017geometric}.
It was further assumed that the scaling behavior of $g(r,L)$ is modified as~\cite{kenna2014fisher}
\begin{equation}\label{g13}
g(r,L) \asymp  r^{-(d-2+\eta_Q)} \tilde{g}(r/L)
\end{equation}
with $\eta_Q=2-d/2$, such that the  decay of $g(r,L)$ is no longer Gaussian-like.
In the study of the 5d Ising model~\cite{luijten1997interaction}, a more subtle scenario was proposed that $g(r,L)$ decays as $r^{-3}$ at short distance,
gradually becomes $r^{-5/2}$ for large distance and has a crossover behavior in between.
The introduction of $\eta_Q$ was refuted by \cite{Wittmann2014} as the magnetic fluctuations at non-zero Fourier mode ${\bf k} \! \neq \!  0$ scale as $\chi_{\bf k} \asymp L^2$ and underlined in~\cite{Flores-Sola} which revealed that the non-zero Fourier moments are
governed by the Gaussian fixed point instead of being contaminated by the dangerously irrelevant field.

Using random-current and random-path representations~\cite{aizenman1982geometric,aizenman1985rigorous,aizenman1986rigorous},
Papathanakos conjectured that the scaling behavior of $g(r,L)$ has a two-length form as~\cite{papathanakos2006finite}
\begin{equation}\label{tp0}
g(r,L) \asymp \begin{cases} r^{-(d-2)}, & r \le \scro \left(L^{d/[2(d-2)]}\right) \\
L^{-d/2},  & r \ge  \scro \left(L^{d/[2(d-2)]}\right) \; .
\end{cases}
\end{equation}
According to~(\ref{tp0}), the critical correlation function still decays as Gaussian-like, $g(r,L) \asymp r^{-(d-2)}$,
up to a length scale $\xi_1=L^{d/[2(d-2)]}$,
and then enters an $r$-independent plateau whose height vanishes as $L^{-d/2}$.
Since the length $\xi_1$ is vanishingly small compared to the linear size, $\xi_1/L \rightarrow 0$,
the plateau effectively dominates the scaling behavior of $g(r,L)$ and the FSS of $\chi$.
The two-length scaling form~(\ref{tp0}) has been numerically confirmed for the 5d Ising model and self-avoiding random walk,
with a geometric explanation based on the introduction of unwrapped length on torus~\cite{grimm2017geometric}.
It is also consistent with the rigorous calculations for the so-called random-length random-walk model~\cite{zhou2018random}.
It is noteworthy that the two-length scaling is able to explain both the FSS $\chi_0 \equiv \chi \asymp L^{5/2}$ for the susceptibility (the magnetic fluctuations at zero Fourier mode)~\cite{luijten1999finite}
and the FSS $\chi_{\bf k} \asymp L^{2}$ for the magnetic fluctuations at non-zero modes~\cite{Wittmann2014,Flores-Sola}.

Combining all the existing numerical and (semi-)analytical insights~\cite{luijten1997interaction,luijten1999finite,papathanakos2006finite,Wittmann2014,kenna2014fisher,Flores-Sola,grimm2017geometric,zhou2018random},
one of us and coworkers extended the scaling (\ref{eq:f_at_dc23}) of the free energy to be~\cite{fang2019}
\begin{equation}\label{eq:f_at_dc234}
\begin{split}
f(t,h) = L^{-d} \tilde{f}_0 (t L^{y_t}, hL^{y_h})+L^{-d} \tilde{f}_1 (t L^{y^*_t}, hL^{y^*_h}) \; ,
\end{split}
\end{equation}
where $(y_t,y_h)$ are the Gaussian exponents~(\ref{eq:gaussian})
and $(y_t^{*},y_h^{*})$ are still given by~(\ref{ystar}).
Conceptually, scaling formula (\ref{eq:f_at_dc234}) explicitly points out the coexistence of two sets of
exponents $(y_t,y_h)$ and $(y_t^{*},y_h^{*})$,
which was implied in previous studies~\cite{Wittmann2014,Flores-Sola,grimm2017geometric,zhou2018random}.
Moreover, a simple perspective of understanding was provided~\cite{fang2019}
that the scaling term with $\tilde{f}_1$ can be regarded to correspond to the FSS of the critical O($n$) model
on a finite complete graph with the number of vertices $V=L^d$.
As a consequence, the exponents $(y_t^{*},y_h^{*})$ can be directly obtained from exact calculations of the complete-graph O($n$) vector model, which also gives $y_t^{*}=d/2$ and $y_h^{*}=3d/4$.
From this correspondence, the plateau of $g(r,L)$ in~(\ref{tp0}) is in line with the FSS of the complete-graph correlation
 function $g_{i\neq j} \equiv \langle \vec{S}_i \cdot \vec{S}_j \rangle $, which also decays as $V^{-1/2} = L^{-d/2}$.
Note that, as a counterpart of the complete-graph scaling function,  the term with $\tilde{f}_1$  should not describe
the FSS of quantities merely associated with $r$-dependent behaviors, including magnetic/energy-like fluctuations at non-zero Fourier modes.
Therefore, in comparison with~(\ref{eq:f_at_dc23}), the scaling formula (\ref{eq:f_at_dc234}) can give the FSS of a more exhaustive list of physical quantities.
The following gives some examples at criticality. %At criticality,
\begin{itemize}
\item let $\vec{\cal M} \! \equiv \! \sum_{\bf r} \vec{S}_{\bf r} $ specify  the total magnetization of a spin configuration,
         and measure its $\ell$ moment as $M_{\ell} \equiv  \langle |\vec{\cal M}|^\ell\rangle $.
         Equation (\ref{eq:f_at_dc234}) predicts $ M_{\ell} \sim L^{\ell y^*_h}+q L^{\ell y_h} $, with $q$ a non-universal constant. In particular, the magnetic susceptibility $\chi_0 \equiv L^{-d} M_2 \asymp L^{d/2} [1+\scro(L^{(4-d)/2})]$,
         where the FSS from the Gaussian term $\tilde{f}_0$ is effectively a finite-size correction
         but its existence is important in analyzing numerical data~\cite{fang2019}.
\item let $\vec{\cal M}_{\bf k}\! \equiv \! \sum_{\bf r} \vec{S}_{\bf r} e^{i {\bf k} \cdot {\bf r}}$
         specify the Fourier mode of magnetization with momentum ${\bf k} \neq 0$,
         and measure its $\ell$ moment as $M_{\ell, {\bf k}} \equiv \langle |\vec{\cal M}_{\bf k} |^\ell \rangle$.
         The magnetic fluctuations at ${\bf k} \neq 0$
         behaves as $ \chi_{\bf k} \equiv L^{-d} M_{2, {\bf k} } \sim L^{2y_h-d}=L^2$.
         The behaviors of $\chi_0$ and $\chi_{\bf k}$ have been confirmed for the 5d Ising model~\cite{Wittmann2014,Flores-Sola,grimm2017geometric,zhou2018random}.
\item the Binder cumulant $Q \equiv  \langle |\vec{\cal M} |^2 \rangle^2/\langle |\vec{\cal M} |^4 \rangle$ should take the complete-graph value, as expected from the correspondence between the term with $\tilde{f}_1$ in (\ref{eq:f_at_dc234}) and the complete-graph FSS.
         For the Ising model, the complete-graph calculations give $Q=4[\frac{\Gamma(3/4)}{ \Gamma(1/4)}]^2 \approx 0.456\,947$,
          consistent with the 5d result in  Ref.~\cite{luijten1997interaction}.
\end{itemize}
Analogously, the FSS behaviors of energy density, its higher-order fluctuations and the $\ell$-moment Fourier modes at ${\bf  k} \neq 0$ can be derived from (\ref{eq:f_at_dc234}).

We expect that the FSS formulae (\ref{tp0}) and (\ref{eq:f_at_dc234}) are valid not only for the O($n$) vector model but also for generic systems of
continuous phase transitions at $d > d_c$.
An example is given for percolation that has $d_c=6$. It was observed~\cite{PhysRevE.97.022107}
that at criticality, the probability distributions of the largest-cluster size follow the same scaling function
for 7d periodic hypercubes and on the complete graph.

In this work, we focus on the FSS for the O($n$) vector model at the upper critical dimensionality $d=d_c$. In this marginal case, it is known that multiplicative and additive logarithmic corrections would appear in the FSS. However, exploring these logarithmic corrections turns out to be of notorious hardness. The challenge comes
  from the lack of analytical insights, the existence of slow finite-size corrections, as well as
   the unavailability of very large system sizes in simulations of high-dimensional systems.

For the O($n$) vector model, establishing the precise FSS form at $d=d_c$ is not only of fundamental importance in statistical mechanics and condensed-matter physics, but also of practical relevance due to the direct experimental realizations of the model,
particularly in three-dimensional quantum critical systems~\cite{Greiner,Capogrosso-Sansone,merchant2014quantum,Qin2015,lohofer2017excitation,qin2017amplitude}. For instance, to explore the stability of Anderson-Higgs excitation modes in systems with continuous symmetry breaking ($n \geq 2$),
a crucial theoretical question is whether or not the Gaussian $r$-dependent behavior $g(r) \asymp  r^{-2}$ is modified by some multiplicative logarithmic corrections.

\section{Summary of main findings}

At the upper critical dimensionality ($d_c=4$) of the O($n$) model,
the state-of-the-art applications of FSS are mostly restricted to a phenomenological scaling form proposed by Kenna~\cite{Kenna2004}
for the singular part of free energy density, which was extended from Aktekin's formula for the Ising model~\cite{Aktekin2001},
\begin{equation}~\label{sc1}
f(t,h)=L^{-4} \tilde{f} (t L^{y_t} ({\rm ln}L)^{\hat{y}_t},hL^{y_h} ({\rm ln}L)^{\hat{y}_h})
\end{equation}
for $n\geq 0$ and $n \neq 4$,
where the renormalization exponents $y_t=2$ and $y_h=3$ are given by (\ref{eq:gaussian}).
Further, the renormalization-group calculations predicted the logarithmic-correction exponents as $\hat{y}_t=\frac{4-n}{2n+16}$ and $\hat{y}_h=1/4$~\cite{wegner1973logarithmic,Kenna2012Chapter}. The leading FSS of $\chi_0$ is hence given by $\chi_0 \asymp L^{2}({\rm ln}L)^{1/2}$, independent of $n$.

Motivated by the recent progress for the O($n$) models for $d>d_c$~\cite{papathanakos2006finite,Wittmann2014,kenna2014fisher,Flores-Sola,grimm2017geometric,zhou2018random,fang2019}, we hereby
propose that at $d=d_c$, the scaling form (\ref{sc1}) for free energy should be revised as
\begin{equation}\label{eq:f_at_dc}
\begin{split}
f(t,h) = & L^{-4} \tilde{f}_0 (t L^{y_t}, hL^{y_h})  +  \\ &  L^{-4} \tilde{f}_1 (t L^{y_t} (\ln L)^{\hat{y}_t}, hL^{y_h} (\ln L)^{\hat{y}_h} ) \; ,
\end{split}
\end{equation}
and the critical two-point correlation $g(r,L)$ behaves as
\begin{equation}\label{tp}
g(r,L) \asymp \begin{cases} r^{-2}, & r \le  \scro \left(L/({\rm ln}L)^{\hat{p}} \right) \\
L^{-2}({\rm ln}L)^{\hat{p}},  & r \ge   \scro \left(L/({\rm ln}L)^{\hat{p}}\right) \; ,
\end{cases}
\end{equation}
with $\hat{p} \! = \! 2 \hat{y}_h \! = \! 1/2$.
By~(\ref{tp}), we explicitly point out that {\it no} multiplicative logarithmic correction appears in
the $r$-dependence of $g(r,L) \! \asymp  \! r^{-2}$, which is still Gaussian-like.
By contrast, the  plateau for $r \! \geq \! \xi_1 \! \sim \! L/({\rm ln}L)^{\hat{p}} $ is modified as $L^{-2}({\rm ln}L)^{\hat{p}}$.
In other words, along any direction of the periodic hypercube,
we have $g(r,L) \! \asymp \! r^{-2} \! + \! v L^{-2}({\rm ln}L)^{\hat{p}}$, with $v$ a non-universal constant.
The decaying with $r^{-2}$ at shorter distance in (\ref{tp}) is consistent with analytical calculations for the 4d weakly self-avoiding random walk and O($n$) $\phi^4$ model directly in the thermodynamic limit ($L \!  \rightarrow \! \infty$)~\cite{slade2016critical}, which predict $g(r) \asymp r^{-2} (1\! + \! \scro(1/{\rm ln} r))$.

The roles of terms with $\tilde{f}_0$ and $\tilde{f}_1$ in~(\ref{eq:f_at_dc}) are analogous to those in (\ref{eq:f_at_dc234}).
The former arises from the Gaussian fixed point, and the latter describes the ``background" contributions (${\bf k}=0$)
for the FSS of macroscopic quantities.
However, it is noted that the term with $\tilde{f}_1$ can no longer be regarded as an exact counterpart of the FSS of complete graph,
due to the existence of multiplicative logarithmic corrections. By contrast, exact complete-graph mechanism applies to the $\tilde{f}_1$ term in (\ref{eq:f_at_dc234}), where logarithmic correction is absent and $\tilde{f}_1$ corresponds to the free energy of standard complete-graph model. According to ~(\ref{eq:f_at_dc}), the FSS of various macroscopic quantities at $d=d_c$ can be obtained as
\begin{itemize}
\item the magnetization density $m \equiv L^{-d} \langle |\vec{\cal M}|\rangle \asymp L^{-1} (\ln L)^{\hat{y}_h} [1+\scro((\ln L)^{-\hat{y}_h}) ]$.
\item the magnetic susceptibility $\chi_0  \asymp L^{2}(\ln L)^{2 \hat{y}_h} [1+\scro((\ln L)^{-2\hat{y}_h}) ] $.
\item the magnetic fluctuations at ${\bf k} \! \neq \!  0$ Fourier modes $\chi_{\bf k} \asymp L^{2}$.
\item the Binder cumulant $Q$ may not take the exact complete-graph value,
         due to the multiplicative logarithmic correction. Some evidence was observed in a recent study by one of us (Y.D.) and his coworkers for the self-avoiding random walk ($n \! = \! 0$) on 4d periodic hypercubes, in which the maximum system size is up to $L=700$.
\end{itemize}
The FSS of energy density, its higher-order fluctuations and the $\ell$-moment Fourier modes at ${\bf  k} \neq 0$ can be obtained.

In quantities like $m$ and $\chi_0$, the FSS from the Gaussian fixed point effectively plays a role as finite-size corrections.
Nevertheless, it is mentioned that in the analysis of numerical data, it is important to include such scaling terms.

We remark that the FSS formulae~(\ref{eq:f_at_dc}) and (\ref{tp}) for $d=d_c$ are less generic than (\ref{tp0}) and (\ref{eq:f_at_dc234}) for $d >d_c$. For the O($n$) models, \textcolor{red}{a} multiplicative logarithmic correction is absent in the Gaussian $r$-dependence of $g(r,L)$ in (\ref{tp}). Albeit the two length scales is possibly a generic feature for models with logarithmic finite-size corrections at upper critical dimensionality, multiplicative logarithmic corrections to the $r$-dependence of $g(r,L)$ require case-by-case analyses. Formula (\ref{eq:f_at_dc}) can be modified in some of these models, which include the percolation and spin-glass models in six dimensions.

We proceed to verify (\ref{eq:f_at_dc}) and (\ref{tp}) using extensive Monte Carlo (MC) simulations of the O($n$) vector model.
Before giving technical details, we present in Fig.~\ref{fig1} complementary evidence
for (\ref{eq:f_at_dc}) and (\ref{tp}) in the example of critical 4d XY model.
Figure~\ref{fig1}(a) shows the extensive data of $g(r,L)$ for $16 \! \leq \! L \! \leq \! 80$,
of which the largest system contains about $4 \times 10^7$ lattice sites.
To demonstrate the multiplicative logarithmic correction in the large-distance plateau indicated by (\ref{tp}),
we plot $g(L/2,L)L^2$ versus ${\rm ln}L$ in the log-log scale in Fig.~\ref{fig1}(b).
The excellent agreement of the MC data with formula $v_1({\rm ln}L)^{1/2} \! + \! v_2$
provides a first-piece evidence for the presence of the logarithmic correction with exponent $\hat{p} \! = \! 1/2$.
The second-piece evidence comes from Fig.~\ref{fig1}(c), suggesting that the $\chi_0 L^{-2}$ data
can be well described by formula $q_1({\rm ln}L)^{1/2} \! + \! q_2$.
Finally, Fig.~\ref{fig1}(d) plots the ${\bf k} \! \neq \!  0$ magnetic fluctuations $\chi_1$ and $\chi_2$ with ${\mathbf k}_1 \! = \!  (2\pi/L, 0, 0, 0)$ and ${\mathbf k}_2 \! = \! (2\pi/L, 2\pi/L, 0, 0)$ respectively, which suppress the $L$-dependent plateau and show the $r$-dependent behavior of $g(r,L)$. Indeed, the $\chi_1 L^{-2}$ and $\chi_2 L^{-2}$ data converge rapidly to constants as $L$ increases.

\section{Numerical results and finite-size scaling analyses}

Using a cluster MC algorithm~\cite{Wolff1989}, we simulate Hamiltonian (\ref{HamOn}) on 4d hypercubic lattices up to $L_{\rm max} \! = \! 96$ (Ising, XY) and $56$ (Heisenberg), and measure a variety of macroscopic quantities including the magnetization density $m$, the susceptibility $\chi_0$, the magnetic fluctuations $\chi_1$ and $\chi_2$, and the Binder cumulant $Q$. Moreover, we compute the two-point correlation function $g(r,L)$ for the XY model up to $L_{\rm max} \! = \! 80$ by means of a state-of-the-art worm MC algorithm~\cite{prokof2001worm}.

\subsection{Estimates of critical temperatures}
In order to locate the critical temperatures $T_c$, we perform least-squares fits for the finite-size MC data of the Binder cumulant to
\begin{equation}~\label{fit1}
Q(L,T)=Q_c+atL^{y_t}({\rm ln}L)^{\hat{y}_t}+b({\rm ln} L)^{-\hat{p}}+c \frac{{\rm ln}({\rm ln}L)}{{\rm ln}L},
\end{equation}
where $t$ is explicitly defined as $T_c-T$, $Q_c$ is a universal ratio, and $a$, $b$, $c$ are non-universal parameters. In addition to the leading additive logarithmic correction, we include $c \frac{{\rm ln}({\rm ln}L)}{{\rm ln}L}$ proposed by~\cite{Kenna2004} as a high-order correction, ensuring the stability of fits. In all fits, we justify the confidence by a standard manner:
the fits with Chi squared $\chi^2$ per degree of freedom (DF) is $\scro(1)$ and remains stable as the cut-off size $L_{\rm min}$ increases. The latter is for a caution against possible high-order corrections not included. The details of the fits are presented in the Supplemental Material (SM).

 \begin{figure}
\includegraphics[width=8cm,height=8cm]{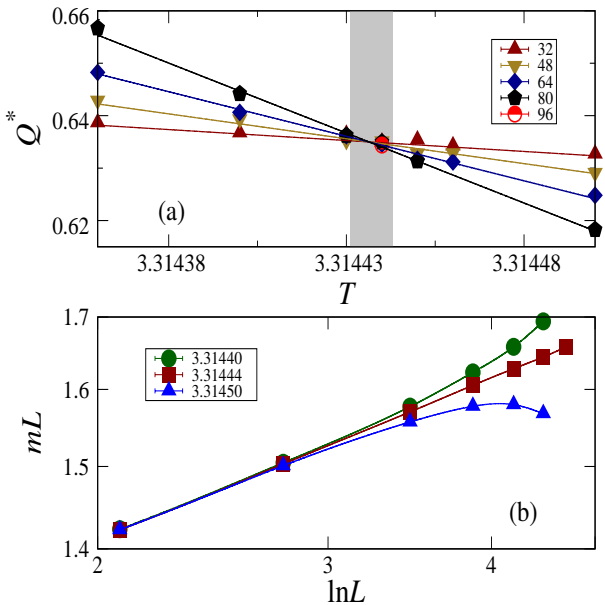}
\caption{Locating $T_c$ for the 4d XY model. (a) The Binder cumulant $Q$ with finite-size corrections being subtracted, namely $Q^*(L,T)=Q(L,T)-b({\rm ln}L)^{-\frac{1}{2}}$, with $b \approx 0.1069$ according to a preferred least-squares fit. The shadow marks $T_c$ and its error margin. (b) The magnetization density $m$ rescaled by $L^{-1}$ versus ${\rm ln}L$ around $T_c=3.314\,44$ in a log-log scale.}~\label{fig2}
\end{figure}

By analyzing the finite-size correction $Q(L,T_c)-Q_c$, we find that the leading correction
is nearly proportional to $({\rm ln} L)^{-1/2}$, consistent with the prediction of (\ref{eq:f_at_dc}) and (\ref{tp}).
We let $Q_c$ be free in the fits and have $Q_c=0.45(1)$, close to the complete-graph result $Q_c=0.456\,947$.
Besides, we perform simulations for the XY and Heisenberg models on the complete graph and obtain as $Q_c\approx 0.635$ and $0.728$, respectively, also close to the fitting results of the 4d $Q$ data.  We obtain $T_c ({\rm XY})=3.314\,437(6)$, and Fig.~\ref{fig2}(a) illustrates the location of $T_c$ by $Q$.

\begin{figure}
\includegraphics[width=8.5cm,height=7cm]{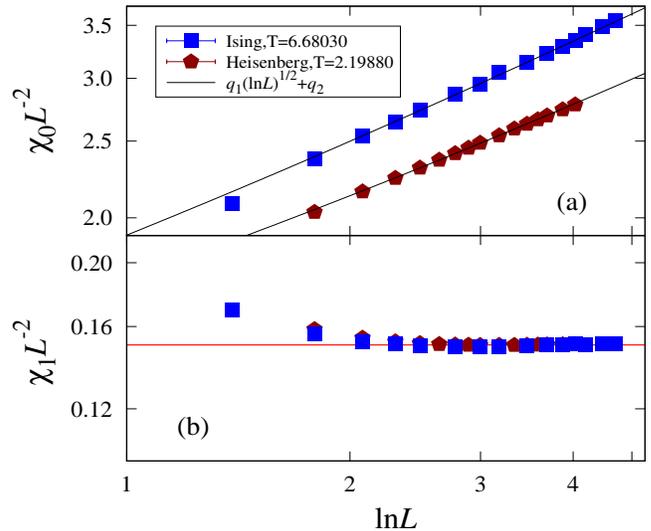}
\caption{The magnetic fluctuations $\chi_0$ (a) and $\chi_1$ (b) rescaled by $L^{2}$ versus ${\rm ln} L$ in a log-log scale for the critical Ising and Heisenberg models. The black lines in (a) represent the least-squares fits, and the red one in (b) denotes a constant.}~\label{fig3}
\end{figure}

We further examine the estimate of $T_c$ by the FSS of other quantities such as the magnetization density $m$.
For the XY model, Fig.~\ref{fig2}(b) gives a log-log plot of the $mL$ data versus ${\rm ln}L$ for
$T=T_c$, as well as for  $T_{\rm low}=3.314\, 40$ and $T_{\rm above}=3.314\, 50$.
The significant bending-up and -down feature clearly suggests that $T_{\rm low} <T_c$ and $T_{\rm above} >T_c$,
providing confidence for the finally quoted error margin of $T_c$.

The final estimates of $T_c$ are summarized in Table~\ref{Tc_summary}. For $ n\! = \! 1$, we have $T_c=6.680\,300(10)$, which is consonant with and improves over $T_c=6.680\,263(23)$~\cite{lundow2009critical} and marginally agrees with $T_c=6.679\,63(36)$~\cite{kenna1991finite} and $6.680\,339(14)$~\cite{luijten1997interaction}. For $ n\! = \! 2$, our determination $T_c=3.314\,437(6)$ significantly improves over $T_c=3.31$~\cite{Jensen2000A,Jensen2000B} and $3.314$~\cite{Nonomura2015}. For $ n\! = \! 3$, our result $T_c=2.198\,79(2)$ rules out $T_c=2.192(1)$ from a high-temperature expansion~\cite{mckenzie1982high}.

\begin{table}
\vspace{-5mm}
 \begin{center}
\caption{Estimates of $T_c$ for the 4d O($n$) vector models.}
 \label{Tc_summary}
 \begin{tabular}[t]{lll}
  \hline
    \hline
   Model &$T_c$& Ref.  \\
 \hline
  {\multirow{4}{*}{Ising ($n=1$)}}
  & $6.679\,63(36)$ & ~\cite{kenna1991finite}  \\
  & $6.680\,339(14)$ & ~\cite{luijten1997interaction} \\
  & $6.680\,263(23)$ &~\cite{lundow2009critical} \\
  & $6.680\,300(10)$ & this work  \\
   \hline
  {\multirow{2}{*}{XY ($n=2$)}}
  &$3.31$, \; $3.314$  &~\cite{Jensen2000A,Jensen2000B,Nonomura2015} \\
 & $3.314\,437(6)$ & this work  \\
   \hline
  {\multirow{2}{*}{Heisenberg ($n=3$) }}
   & $2.192(1)$  & ~\cite{mckenzie1982high}  \\
   & $2.198\,79(2)$  & this work  \\
 \hline
  \hline
 \end{tabular}
 \end{center}
 \vspace{-5mm}
 \end{table}

\subsection{Finite-size scaling of the two-point correlation}

We then fit the critical two-point correlation $g(L/2,L)$ to
\begin{equation}~\label{sccgrL}
g(L/2,L) = v_1 L^{-2}({\rm ln}L)^{\hat{p}} + v_2 L^{-2},
\end{equation}
where the first term comes from the large-distance plateau and the second one
is from the $r$-dependent behavior of $g(r,L)$. With $\hat{p}=1/2$ being fixed,
the estimate of leading scaling term $L^{-1.98(4)}$ agrees well with the exact $L^{-2}$.
With the exponent $-2$ in $L^{-2}$ being fixed, the result $\hat{p}=0.5(1)$ is also well consistent with the prediction $\hat{p}=1/2$. These results are elaborated in the SM.

We remark that FSS analyses for $g(L/2,L)$ have already been performed in~\cite{kenna2014fisher} with the formula $g(L/2,L)=A L^{-2}[{\rm ln} (L/2+B)]^{1/2}$ ($A$ and $B$ are constants) and in~\cite{luijten1997interaction} with a similar formula. These FSS in literature correspond to the first scaling term in Eq.~(\ref{sccgrL}). Hence, formula (\ref{sccgrL}) serves as a forward step for complete FSS by involving the scaling term $v_2 L^{-2}$, which arises from the Gaussian fixed point.

\subsection{Finite-size scaling of the magnetic susceptibility}
According to (\ref{eq:f_at_dc}) and (\ref{tp}), we fit the critical susceptibility $\chi_0$ to
\begin{equation}~\label{scchi0}
\chi_0 = q_1 L^{2}({\rm ln}L)^{\hat{p}}+ q_2 L^{2},
\end{equation}
with $q_1$ and $q_2$ non-universal constants. For $\hat{p}=1/2$ being fixed,
we obtain fitting results with $\chi^2 /{\rm DF} \lesssim  1$ for each of $ n \! = \! 1,2,3$,
and correctly produce the leading scaling form $L^2$. The scaled susceptibility $\chi_0 L^{-2}$ versus ${\rm ln}L$ are demonstrated by Figs.~\ref{fig1}(c) (XY) and~\ref{fig3}(a) (Ising and Heisenberg).

We note that previous studies based on a FSS without high-order corrections produced estimates of ${\hat y}_h$ $(=\hat{p}/2)$, considered to be consistent with ${\hat y}_h=1/4$~\cite{lai1990finite,kenna1991finite,kenna1993renormalization,kenna1994scaling}.
The maximum lattice size therein was $L_{\rm max}=24$, four times smaller than $L_{\rm max}=96$ of the present study. In particular, it was reported~\cite{lai1990finite} that $2\hat{y}_h=0.45(8)$ and $4\hat{y}_h=0.80(25)$.
Nevertheless, we find that the fit $\chi_0= q_1 L^2 ({\rm ln}L)^{2 \hat{y}_h}$ by dropping the correction term $q_2 L^2$
would yield $\hat{y}_h=0.21(1)$ (Ising), $0.20(1)$ (XY), and $0.19(1)$ (Heisenberg),
which are smaller than and inconsistent with the predicted value $\hat{y}_h=1/4$.
This suggests the significance of $q_2 L^2$ in the susceptibility $\chi_0$, which arises from the $r$-dependence of $g(r,L)$.

\begin{figure}
\includegraphics[width=8.5cm,height=7cm]{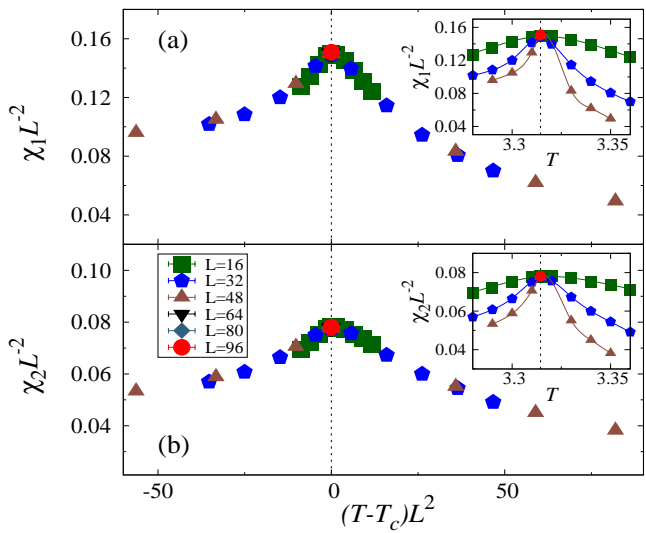}
\caption{Data collapses for the magnetic fluctuations $\chi_1$ (a) and $\chi_2$ (b) rescaled by $L^{2 y_h-d}$ and $L^{y_t}$ ($y_h=3$, $y_t=2$, $d=4$) for the 4d XY model. The insets show the scaled fluctuations versus $T$, and the dashed lines denote $T_c$.}~\label{fig4}
\end{figure}

\subsection{Finite-size scaling of the magnetic fluctuations at non-zero Fourier modes}

We consider the magnetic fluctuations $\chi_1$ with $|{\bf k}_1|= 2 \pi /L$ and $\chi_2 $ with $|{\bf k}_2|= 2 \sqrt{2} \pi /L$.
We have compared the FSS of $\chi_0$, $\chi_1$ and $\chi_2$ in Figs.~\ref{fig1}(c) and (d) for the critical 4d XY model.
As $L$ increases, $\chi_1 L^{-2}$ and $\chi_2 L^{-2}$ converge rapidly, suggesting the absence
of multiplicative logarithmic correction. This is in sharp contrast to the behavior of $\chi_{0} L^{-2}$, which diverges logarithmically. For the Ising and Heisenberg models, the FSS of the fluctuations at non-zero modes is also free of multiplicative logarithmic correction (Fig.~\ref{fig3}(b)).

Surprisingly, it is found that the scaled fluctuations $\chi_{1} L^{-2} \approx 0.15 $ are equal within error bars for the
Ising, XY, and Heisenberg models.

Further, we show in Fig.~\ref{fig4} $\chi_{1}$ and $\chi_{2}$ versus $T$ for the 4d XY model.
It is observed that the magnetic fluctuations at non-zero Fourier modes reach maximum at $T_c$
and that the $\chi_1 L^{-2}$ ($\chi_2L^{-2}$) data for different $L$s collapse well not only at $T_c$ but also
for a wide range of $(T-T_c)L^{y_t}$ with $y_t=2$.

\section{Discussions}
We propose formulae (\ref{eq:f_at_dc}) and (\ref{tp}) for
the FSS of the O($n$) universality class at the upper critical dimensionality, which are tested against extensive MC simulations with $n \! = \! 1, 2, 3$. From the FSS of the magnetic fluctuations at zero and non-zero Fourier modes, the two-point correlation function, and the Binder cumulant, we obtain complementary and solid evidence supporting (\ref{eq:f_at_dc}) and (\ref{tp}). As byproducts, the critical temperatures for $n \! = \! 1,2,3$ are all located up to an unprecedented precision.

An immediate application of (\ref{tp}) is to the massive amplitude excitation mode (often called the Anderson-Higgs boson) due to the spontaneous breaking of the continuous O($n$) symmetry~\cite{pekker2015amplitude}, which is at the frontier of condensed matter research.
At the pressure-induced quantum critical point (QCP) in the dimerized quantum antiferromagnet TlCuCl$_{\rm 3}$,
the 3D O(3) amplitude mode was probed by neutron spectroscopy and
a rather narrow peak width of about 15\% of the excitation energy was revealed,
giving no evidence for the logarithmic reduction of the width-mass ratio~\cite{merchant2014quantum}.
This was later confirmed by quantum MC study of a 3D model Hamiltonian of O(3) symmetry~\cite{lohofer2017excitation,qin2017amplitude}.
Indeed, (\ref{tp}) provides an explanation why the logarithmic-correction reduction in the Higgs resonance was not observed at 3D QCP.
In numerical studies of the Higgs excitation mode at 3D QCP,
the correlation function $g(\tau  \! \equiv \!  |\tau_1 \! - \! \tau_2|)$ is measured along the imaginary-time axis $\beta$,
and numerical analytical continuation is used to deal with the $g(\tau)$ data.
In practice, simulations are carried out at very low temperature $\beta \! \rightarrow \! \infty$,
and it is expected that $g(\tau)  \! \asymp \! \tau^{-2}$ for a significantly wide range of $\tau$.
Furthermore, it is the $\tau$-dependent behavior of $g(\tau)$, instead of the $L$-dependence, that plays a decisive role in numerical analytical continuation.

In the thermodynamic limit, the two-point correlation function decays as $g(r) \sim r^{-2} \tilde{g}(r/\xi)$,
where the scaling function $\tilde{g}(r/\xi)$ quickly drops to zero as $r/\xi \gg 1$. It can be seen that no multiplicative logarithmic correction exists in the algebraic decaying behavior.
On the other hand, as the criticality is approached ($t \! \rightarrow \! 0$),
the correlation length diverges as $\xi(t) \! \sim \!  t^{-1/2}|{\rm ln}t|^{\hat{\nu}}$, and $\hat{\nu} \! = \! \frac{n+2}{2(n+8)} \! > \! 0$ implies that $\xi$ diverges faster than $t^{-1/2}$~\cite{Kenna2004,Kenna2012Chapter}.
Since the susceptibility can be calculated by summing up the correlation as $\chi_0 \sim \int_0^{\xi} g(r)r^{d-1}dr \sim \xi^2 $,
one has $\chi_0(t) \sim t^{-1} |\ln t|^{\hat{\gamma}}$ with $\hat{\gamma} = 2 \hat{\nu}$. The thermodynamic scaling of $\chi_0(t)$ can also be obtained from the FSS formula~(\ref{sc1}) or (\ref{eq:f_at_dc}), which gives $\chi_0(t,L) \sim L^{2y_h-4} (\ln L)^{2\hat{y}_h} \tilde{\chi_0} (tL^{y_t} (\ln L)^{\hat{y}_t})$. By fixing $tL^{y_t} (\ln L)^{\hat{y}_t}$ at some constant, one obtains the relation $L \sim t^{-1/y_t} |\ln t|^{-\hat{y}_t/y_t}$. Substituting it into the FSS of $\chi_0(t,L)$ yields $\chi_0(t) \sim t^{\gamma} |\ln t|^{\hat{\gamma}}$ with $\gamma \! = \!  (2y_h-4)/y_t$ and $\hat{\gamma} \! = \! -\gamma \hat{y}_t+2\hat{y}_h$. With $(y_t, y_h, \hat{y}_t, \hat{y}_h)=(2, 3, \frac{4-n}{2n+16}, \frac{1}{4})$, one has $\gamma = 1$ and $\hat{\gamma} = \frac{n+2}{n+8}$. The thermodynamic scaling with logarithmic corrections has been demonstrated in Ref.~\cite{Qin2015} in terms of the magnetization $m$ of an O(3) Hamiltonian.

For the critical Ising model in five dimensions, an unwrapped distance $r_{\rm u}$ was introduced to account for the winding numbers across a finite torus~\cite{grimm2017geometric}. The unwrapped correlation was shown to behave as $g(r_{\rm u})  \sim r_{\rm u}^{2-d} \tilde{g} (r_{\rm u}/\xi_{\rm u})$, where the unwrapped correlation length diverges as $\xi_{\rm u} \sim L^{d/4}$. This differs from typical correlation functions that are cut off by the linear system size $\sim L$.
We expect that at $d_c=4$, the unwrapped correlation length diverges as $\xi_{\rm u} \sim L (\ln L)^{\hat{y}_h}$, which gives the critical susceptibility as $\chi_0(L) \sim L^2 (\ln L)^{2\hat{y}_h}$.

Besides, formula (\ref{tp}) is useful for predicting various critical behaviors. As an instance, it was observed that an impurity immersed in a 2D O(2) quantum critical environment can evolve into a quasiparticle of fractionalized charge, as the impurity-environment interaction is tuned to a boundary critical point~\cite{huang2016trapping,whitsitt2017critical,chen2018halon}. Formula (\ref{tp}) precludes the emergence of such a quantum-fluctuation-induced quasiparticle at 3D O(2) QCP.

We mention an open question about the specific heat of the 4d Ising model. FSS formula (\ref{sc1}) predicts that the critical specific heat diverges as $C \! \asymp \! (\ln L)^{1/3}$. By contrast, a MC study demonstrated that the critical specific heat is bounded~\cite{lundow2009critical}. The complete scaling form (\ref{eq:f_at_dc}) is potentially useful for reconciling the inconsistence.

Finally, it would be possible to extend the present scheme to other systems of critical phenomena, as the existence of upper critical dimensionality is a common feature therein. These systems include the percolation and spin-glass models at their upper critical dimensionality $d_c=6$. We leave this for a future study.

\section{Method}
Throughout the paper, the raw data for any temperature $T$ and linear size $L$ are obtained by means of MC simulations, for which the Wolff cluster algorithm~\cite{Wolff1989} and the Prokof'ev-Svistunov worm algorithm~\cite{prokof2001worm} are employed complementarily. Both algorithms are state-of-the-art tools in their own territories.

The O($n$) vector model (\ref{HamOn}) in its original spin representation is efficiently sampled by the Wolff cluster algorithm, which is the single-cluster version of the widely utilized non-local cluster algorithms. The present study uses the standard procedure of the algorithm, as in the original paper~\cite{Wolff1989} where the algorithm was invented. In some situations, we also use the conventional Metropolis algorithm~\cite{metropolis1953equation} for benchmarks. The macroscopic physical quantities of interest have been introduced in aforementioned sections for the spin representation.

The two-point correlation function for the XY model ($n=2$) is sampled by means of the Prokof'ev-Svistunov worm algorithm, which was invented for a variety of classical statistical models~\cite{prokof2001worm}. By means of a high-temperature expansion, we perform an exact transformation for the original XY spin model to a graphic model in directed-flow representation. We then introduce two defects for enlarging the state space of directed flows. The Markov chain process of evolution is built upon biased random walks of defects, which satisfy the detailed balance condition. It is defined that the evolution hits the original directed-flow state space when the two defects meet at a site. The details for the exact transformation and a step-by-step procedure for the algorithm have been presented in a recent reference~\cite{Xu2019}.

\begin{acknowledgments}
\textit{Acknowledgements.} YD is indebted to valuable discussions with Timothy Garoni, Jens Grimm and Zongzheng Zhou.
 This work has been supported by the National Natural Science Foundation of China under Grants No.~11774002, No.~11625522, and No.~11975024, the National Key R\&D Program of China under Grants No.~2016YFA0301604
and No.~2018YFA0306501, and the Department of Education in Anhui Province.
\end{acknowledgments}

\section{Conflict of interest statement}
None declared.

\bibliography{papers}

\end{document}

\author{Jian-Ping Lv}
\email{jplv2014@ahnu.edu.cn}
\author{Wanwan Xu}
\author{Yanan Sun}
\affiliation{Department of Physics, Anhui Key Laboratory of Optoelectric Materials Science and Technology, Key Laboratory of Functional Molecular Solids, Ministry of Education, Anhui Normal University, Wuhu, Anhui 241000, China}
\author{Kun Chen}
\email{chenkun0228@gmail.com}
\affiliation{Department of Physics and Astronomy, Rutgers, The State University of New Jersey, Piscataway, New Jersey 08854-8019}
\author{Youjin Deng}
\email{yjdeng@ustc.edu.cn}
\affiliation{National Laboratory for Physical Sciences at Microscale and Department of Modern Physics, University of Science and Technology of China, Hefei, Anhui 230026, China}
\affiliation{Department of Physics and Electronic Information Engineering, Minjiang University, Fuzhou, Fujian 350108, China}

\title{Supplemental Material for ``Finite-size Scaling of O(n) Systems at the Upper Critical Dimensionality''}

\date{\today}

\maketitle

We elaborate the quantifications of the finite-size scaling (FSS) mentioned in the main text. Subsequently, we address the location of the critical temperatures $T_c$, the FSS of the susceptibility $\chi_0$, and the FSS of the two-point correlation $g(r,L)$ with $r=L/2$.

\section{The estimate of $T_c$}
For each of the four-dimensional Ising, XY, and Heisenberg models, the estimate of $T_c$ is achieved by fitting the finite-size Monte Carlo data of the Binder cumulant $Q$ to the scaling ansatz
\begin{equation}~\label{fit1}
Q(L,T)= Q_c+ a t L^{y_t}({\rm ln}L)^{\hat{y}_t}+ b ({\rm ln} L)^{-\hat{p}}+ c \frac{{\rm ln} ({\rm ln}L)}{{\rm ln}L}
\end{equation}
where $t \equiv T_c-T$, for $t \rightarrow 0$. $Q_c$ is the critical dimensionless ratio and $a$, $b$, $c$ are constants. In the fits, the mean-field thermal exponent is fixed as $y_t=2$ for reducing uncertainty.

For the Ising model, as shown in Table~\ref{fit_result_1_1}, we perform fits with ${\hat{y}_t}$ being fixed to be $\hat{y}_t=\frac{4-n}{2n+16}=1/6$ as predicated by renormalization-group calculations~\cite{Kenna2012Chapter} or being free. After obtaining an estimate of $T_c$, we also perform fits right at $T_c$. Stable fits with $\chi^2/{\rm DF} \lessapprox 1$ are achieved for all of these scenarios. As we let $Q_c$ free, it is found $Q_c=0.45(1)$, which is close to $Q_c=0.456\,947$ of the complete-graph model~\cite{luijten1997interaction}. By these fits, we estimate ${\hat p}\approx 0.5$. The fits for $Q$ of the XY and Heisenberg models are summarized in Tables~\ref{fit_result_1_2} and ~\ref{fit_result_1_3}, where the estimates for $Q_c$ are again close to the results $Q_c \approx 0.635$ and $0.728$ by Monte Carlo simulations of XY and Heisenberg models on complete graph~\cite{Hu2019}, respectively. Meanwhile, we note that the amplitude of correction $b \approx 0.1$ is sizeable for each of the four-dimensional models. Hence, the prediction of this study on the existence of the finite-size correction $b ({\rm ln} L)^{-\hat{p}}$ is further confirmed. This correction form is visualized by Fig.~\ref{SM_fig1} for the Ising, XY, and Heisenberg models. The locating of $T_c$ is shown in Fig.~{\ref{SM_fig2}.

\begin{figure}
\includegraphics[width=8.5cm,height=5cm]{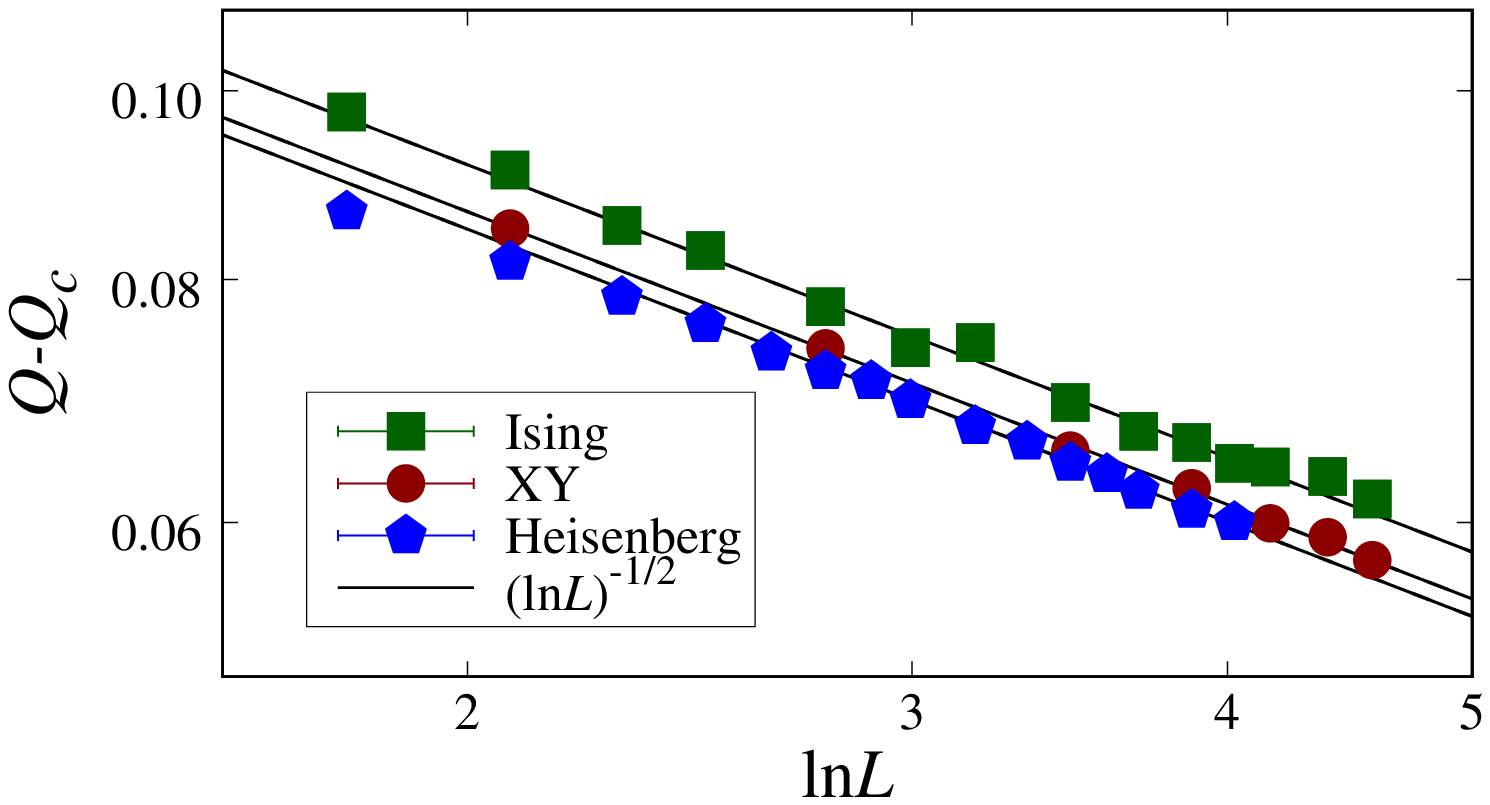}
\caption{Finite-size corrections $Q(L,T_c)-Q_c$ of the Binder cumulant $Q$ versus ${\rm ln} L$ in a log-log scale for the critical Ising, XY, and Heisenberg models. The solid lines are drawn according to preferred fits and stand for the $({\rm ln}L)^{-\frac{1}{2}}$ decaying.}~\label{SM_fig1}
\end{figure}

\begin{figure}
\includegraphics[width=8.5cm,height=12cm]{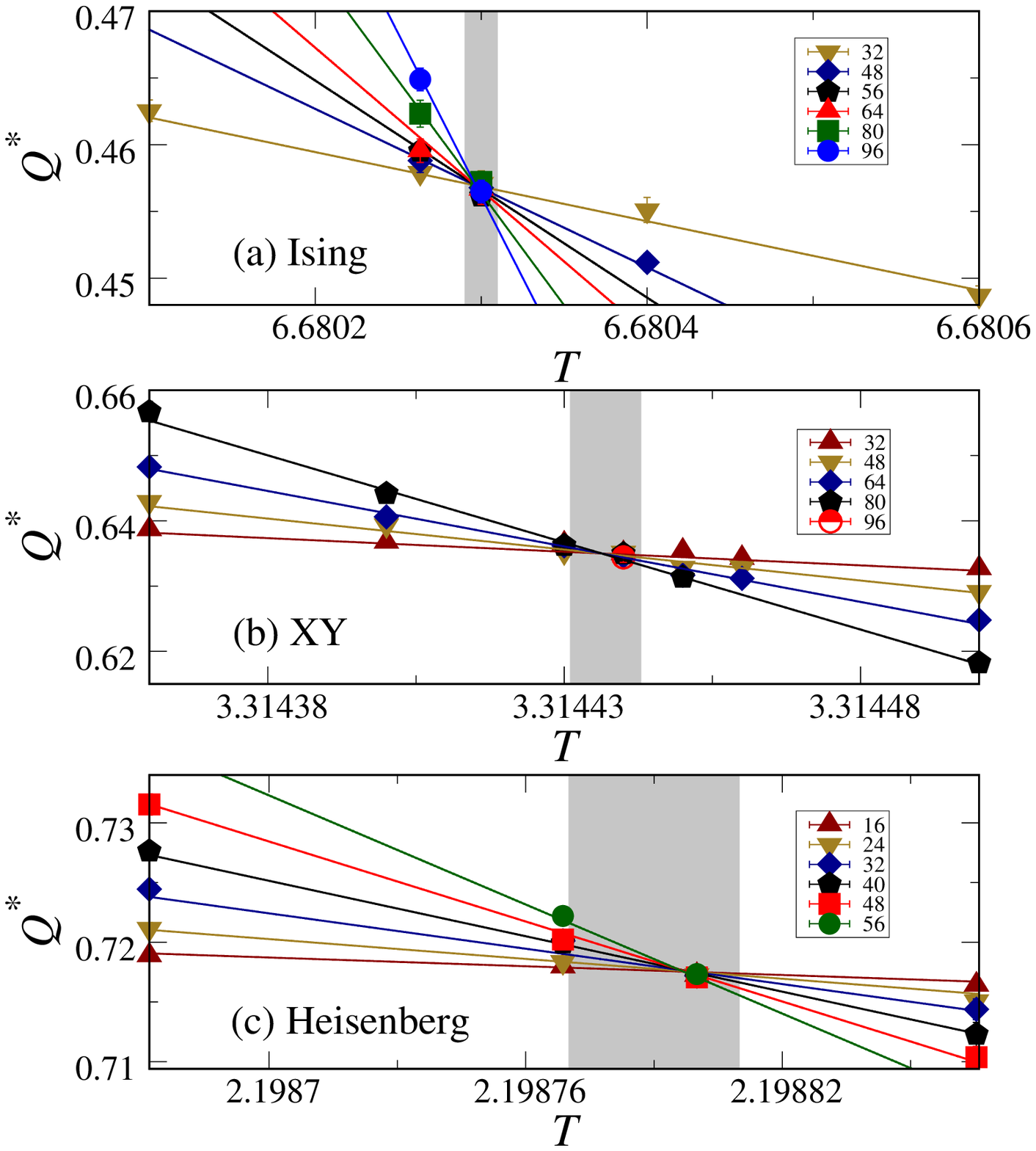}
\caption{The Binder cumulant $Q$ with finite-size corrections being subtracted, namely $Q^*(L,T)=Q(L,T)-b({\rm ln}L)^{-\frac{1}{2}}$, with $b\approx 0.0991 $, $0.1069$, and $0.1160$ according to the preferred least-squares fits for the Ising (a), XY (b), and Heisenberg (c) models, respectively. The solid lines are drawn according to the fits, and the shadows mark $T_c$ and their error margins. The plot for the XY case has appeared in the main text and we hereby include it for completeness.}~\label{SM_fig2}
\end{figure}

\begin{figure}
\includegraphics[width=8.5cm,height=12cm]{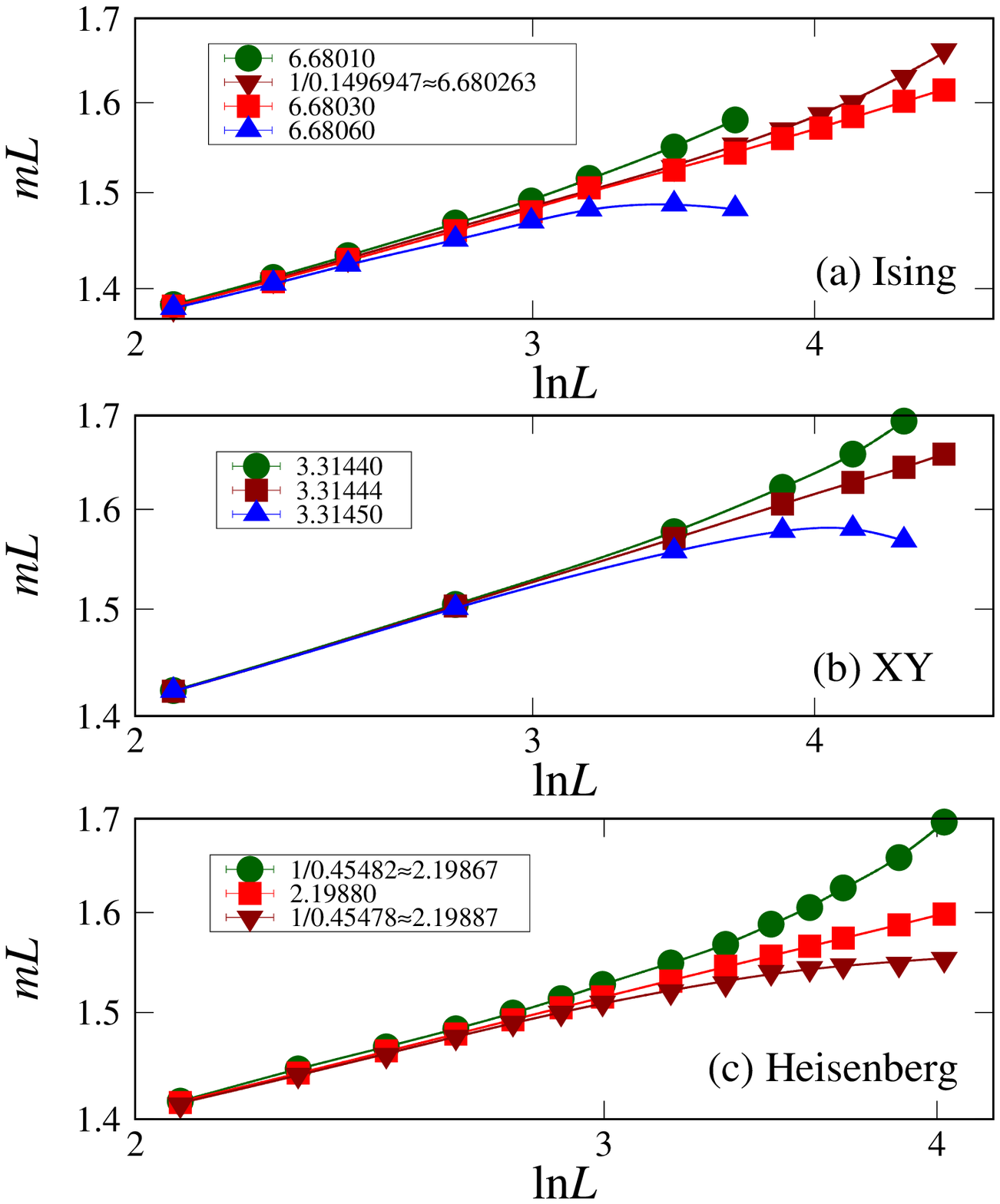}
\caption{The magnetization density $m$ rescaled by its mean-field factor $L^{y_h-d}$ ($y_h=3$, $d=4$) versus ${\rm ln}L$ around $T_c=6.680\,30$, $3.314\,44$, and $2.198\,80$ in a log-log scale, for the Ising (a), XY (b), and Heisenberg (c) models, respectively. For each of the models, linearity is observed at $T_c$, and the bending-up or -down feature with respect to the linearity indicates the deviation from criticality. The plot for the XY case has appeared in the main text and we hereby include it for completeness.}~\label{SM_fig3}
\end{figure}

The final estimates of $T_c$ are determined by comparing the fits, and are $T_c=6.680\,30(1)$, $3.314\,437(6)$, and $2.198\,79(2)$, for the Ising, XY, and Heisenberg models, respectively. These estimates can be examined independently using the quantities other than $Q$, e.g., the magnetization density $m$. Figure~\ref{SM_fig3} displays $m$ rescaled by the mean-field factor $L^{y_h-d}$ ($y_h=3$, $d=4$) versus ${\rm ln}L$ in a log-log scale around $T_c$ for the Ising, XY, and Heisenberg models. The linearities at $T_c$ in the plots confirm that the estimated $T_c$ are reasonable for the present lattice size scale, while the bending-up or -down features with respect to the linearities suggest the deviations from $T_c$. For the Ising model, we confirm the linearity at $T_c=6.680\,30$, and preclude the temperatures $T=6.680\,263$ and $6.680\,60$ as $T_c$. For the XY model, we confirm $T_c=3.314\,44$, and preclude $T_c=3.314\,40$ and $3.314\,50$. For the Heisenberg model, $T_c=2.198\,80$ is confirmed and $T_c=2.198\,67$ and $2.198\,87$ are both precluded.

\section{FSS of the susceptibility $\chi_0$}
We fit the Monte Carlo data of the susceptibility $\chi_0$ at $T_c$ to the ansatz
\begin{equation}~\label{fit2}
\chi_0(L,T_c)= q_1 L^{2 y_h-d}({\rm ln}L)^{\hat{p}}+ q_2 L^{2 y_h-d},
\end{equation}
which is an inference drawn from the scaling formulae of free energy density and two-point correlation. The fits are summarized in Table~\ref{fit_result_2} for the critical Ising, XY, and Heisenberg models. For each of the models, we confirm that, if one includes both $q_1$ and $q_2$ terms and let $\hat{p}=1/2$ and $2 y_h-d=2$ fixed, stable fits with $\chi^2/{\rm DF} \lessapprox 1$ can be achieved. If the mean-field exponent $2y_h-d$ is free in the fits, we obtain $2y_h-d \approx  2.00$ for each model, in perfect agreement with the exact value $2$ within error bars. For the Ising and XY models (for which we have Monte Carlo data with $L_{\rm max}=96$), we achieve stable fits with $\hat{p}$ being free, which yield $\hat{p} \approx 0.5$, consistent with the prediction $\hat{p}=1/2$.

\section{FSS of the two-point correlation $g(L/2,L)$}

We perform FSS for the large-distance plateau of two-point correlation $g(r,L)$, by fitting the finite-size $g(L/2,L)$ data to the ansatz
\begin{equation}\label{fit3}
g(L/2,L) = v_1 L^{2y_h-2d}({\rm ln}L)^{\hat{p}} + v_2 L^{2y_h-2d}
\end{equation}
for the critical XY model. We perform fits under the situations that $2y_h-2d=-2$ and $\hat{p}=1/2$ are both fixed,
and that only one of them is fixed. The fits are summarized by Table~\ref{fit_result_3}, which demonstrates that the inclusion of
both $v_1$ and $v_2$ terms correctly produces the mean-field exponent $y_h$. As an instance, one of the fits yields $2y_h-2d=-1.99(1)$ with $L_{\rm min}=24$ and $\chi^2/{\rm DF} \approx 0.8$, which is in good agreement with the exact $2y_h-2d=-2$. By comparing the fits with $\hat{p}=1/2$ being fixed, we estimate $y_h=3.01(2)$. As a further verification, once the mean-field exponent $2y_h-2d=-2$ is fixed, the estimate $\hat{p}=0.5(1)$ is again consistent with the prediction $\hat{p}=1/2$.

\makeatletter

\makeatother

\newpage

\begin{table*}
 \begin{center}
\caption{Fits of the finite-size Monte Carlo data of the Binder cumulant $Q$ to (\ref{fit1}) for the Ising ($n=1$) model. The missing entries mean that the corresponding high-order corrections are not included or that the fits are performed right at $T_c=6.680\,30$.}
 \label{fit_result_1_1}
 \begin{tabular}[t]{llllllllll}
 \hline
  \hline
    $L_{\rm min}$ \qquad\qquad   & $\chi^2/$DF \qquad \qquad & $T_c$ \qquad  \qquad \qquad\qquad& ${\hat y_t}$ \qquad \qquad \qquad  &${\hat p}$  \qquad \qquad  \qquad & $Q_c$ \qquad \qquad \qquad&$a$ \qquad\qquad \qquad & $b$  \qquad \qquad \qquad & $c$  \qquad \qquad  \\
 \hline
  8& 61.6/45&6.680\,310(6)&$-0.3(2)$&0.5(1)&$0.44(2)$&$0.04(1)$&0.13(1)&\\
  32&9.0/16&6.680\,307(9)&$-0.5(5)$&0.6(1)&$0.456947$&$0.05(3)$&0.12(1)&\\
  32&10.7/17&6.680\,302(6)&$1/6$&0.59(9)&$0.456947$&$0.0205(8)$&0.11(1)&\\
  40&9.2/12&6.680\,302(9)&$1/6$&0.6(2)&$0.456947$&$0.020(1)$&0.11(3)&\\
  24&34.1/22&6.680\,292(2)&$1/6$&0.5&$0.456947$&$0.0207(8)$&0.1004(3)&\\
  32&11.6/18&6.680\,297(3)&$1/6$&0.5&$0.456947$&$0.0206(8)$&0.0991(5)&\\
  40&9.4/13&6.680\,298(3)&$1/6$&0.5&$0.456947$&$0.020(1)$&0.0987(6)&\\
  48&2.0/8&6.680\,301(4)&$1/6$&0.5&$0.456947$&$0.019(1)$&0.098(1)&\\
  56&1.7/5&6.680\,302(5)&$1/6$&0.5&$0.456947$&$0.019(2)$&0.098(1)&\\
  24&18.8/22&6.680\,307(4)&$1/6$&0.5&$0.456947$&$0.0203(8)$&0.17(1)&-0.11(2)\\
  32&10.7/17&6.680\,301(5)&$1/6$&0.5&$0.456947$&$0.0205(8)$&0.13(3)&-0.05(5)\\
  40&9.2/12&6.680\,302(8)&$1/6$&0.5&$0.456947$&$0.020(1)$&0.13(7)&0.0(1)\\
  12&14.8/8&&&0.5&$0.441(2)$&&0.130(3)&\\
  16& 6.2/7&&&0.5&$0.446(3)$&&0.119(5)&\\
  20& 2.8/6&&&0.5&$0.451(4)$&&0.110(7)&\\
  32& 1.8/5&&&0.5&$0.457(7)$&&0.10(1)&\\
  40& 1.2/4&&&0.5&$0.460(8)$&&0.09(2)&\\
   8& 10.1/9&&&0.5&$0.460(5)$&&0.137(2)&-0.06(2)\\
   20& 2.6/6&&&0.56(3)&$0.456947$&&0.107(5)&\\
   32& 1.8/5&&&0.50(7)&$0.456947$&&0.099(9)&\\
   8 & 10.1/9&&&0.48(4)&$0.456947$&&0.140(6)&-0.07(2)\\
 \hline
 \hline
 \end{tabular}
 \end{center}

\begin{center}
\caption{Fits of the finite-size Monte Carlo data of the Binder cumulant $Q$ to (\ref{fit1}) for the XY ($n=2$) model. The missing entries mean that the corresponding high-order corrections are not included or that the fits are performed right at $T_c=3.314\,44$.}
 \label{fit_result_1_2}
 \begin{tabular}[t]{llllllllll}
 \hline
  \hline
    $L_{\rm min}$ \qquad\qquad   & $\chi^2/$DF \qquad \qquad & $T_c$ \qquad  \qquad \qquad\qquad& ${\hat y_t}$ \qquad \qquad \qquad  &${\hat p}$  \qquad \qquad  \qquad & $Q_c$ \qquad \qquad \qquad  &$a$ \qquad\qquad \qquad & $b$  \qquad \qquad \qquad & $c$  \qquad \qquad  \\
 \hline
   8& 29.9/35&3.314\,440(3)&$-0.2(3)$&0.5(1)&$0.62(3)$&$0.05(2)$&0.13(2)&\\
   8& 30.6/36&3.314\,440(3)&$1/10$&0.5(1)&$0.62(3)$&$0.0360(8)$&0.13(2)&\\
   8& 30.7/37&3.314\,439(2)&$1/10$&0.5&$0.63(1)$&$0.0360(8)$&0.126(2)&\\
   16& 25.8/30&3.314\,439(2)&$1/10$&0.5&$0.62(1)$&$0.0360(8)$&0.126(4)&\\
   32& 23.6/23&3.314\,442(4)&$1/10$&0.5&$0.62(1)$&$0.0360(8)$&0.13(1)&\\
   16& 50.4/31&3.314\,430(1)&$1/10$&0.5&$0.635$&$0.0358(8)$&0.1091(2)&\\
   32& 26.9/24&3.314\,434(1)&$1/10$&0.5&$0.635$&$0.0359(8)$&0.1078(3)&\\
   48& 18.1/17&3.314\,436(2)&$1/10$&0.5&$0.635$&$0.0360(8)$&0.1069(7)&\\
   16& 26.4/30&3.314\,437(2)&$1/10$&0.5&$0.635$&$0.0360(8)$&0.131(4)&-0.035(7)\\
   32& 23.7/23&3.314\,440(4)&$1/10$&0.5&$0.635$&$0.0360(8)$&0.15(3)&-0.07(4)\\
   8& 1.9/5&&&0.5&$0.63(1)$&&0.123(2)&\\
   16& 1.9/4&&&0.5&$0.63(1)$&&0.123(4)&\\
   32& 0.8/3&&&0.5&$0.63(1)$&&0.11(1)&\\
    8& 1.9/4&&&0.5(2)&$0.63(2)$&&0.12(2)&\\
   32& 0.8/3&&&0.53(5)&$0.635$&&0.111(7)&\\
   48& 0.6/2&&&0.6(1)&$0.635$&&0.12(2)&\\
    8& 1.8/4&&&0.59(9)&$0.635$&&0.114(8)&0.01(3)\\
   16& 1.0/3&&&0.4(1)&$0.635$&&0.15(4)&-0.07(9)\\
 \hline
 \hline
 \end{tabular}
 \end{center}
 \end{table*}

 \begin{table*}
 \begin{center}
\caption{Fits of the finite-size Monte Carlo data of the Binder cumulant $Q$ to (\ref{fit1}) for the Heisenberg ($n=3$) model. The missing entries mean that the corresponding high-order corrections are not included or that the fits are performed right at $T_c=2.198\,80$.}
 \label{fit_result_1_3}
 \begin{tabular}[t]{lllllllllll}
 \hline
  \hline
    $L_{\rm min}$ \qquad\qquad   & $\chi^2/$DF \qquad \qquad & $T_c$ \qquad  \qquad \qquad\qquad& ${\hat y_t}$ \qquad \qquad \qquad  &${\hat p}$  \qquad \qquad  \qquad & $Q_c$ \qquad \qquad \qquad  &$a$ \qquad\qquad \qquad & $b$  \qquad \qquad \qquad & $c$  \qquad \qquad  \\
   \hline
   24& 22.4/24&2.198\,796(8)&$1/22$&0.59(7)&$0.728$&$0.045(1)$&0.108(9)&\\
   12& 41.0/44&2.198\,791(3)&$1/22$&0.5&$0.72(1)$&$0.045(1)$&0.108(2)&\\
   16& 32.7/36&2.198\,796(4)&$1/22$&0.5&$0.72(1)$&$0.046(1)$&0.114(5)&\\
   20& 25.4/28&2.198\,799(6)&$1/22$&0.5&$0.72(1)$&$0.046(1)$&0.118(8)&\\
   24& 22.4/24&2.198\,797(8)&$1/22$&0.5&$0.72(1)$&$0.045(1)$&0.11(1)&\\
   20& 31.2/29&2.198\,785(2)&$1/22$&0.5&$0.728$&$0.045(1)$&0.0981(2)&\\
   24& 23.9/25&2.198\,788(2)&$1/22$&0.5&$0.728$&$0.045(1)$&0.0976(3)&\\
   32& 20.9/17&2.198\,789(3)&$1/22$&0.5&$0.728$&$0.045(1)$&0.0974(5)&\\
   12& 42.3/44&2.198\,788(3)&$1/22$&0.5&$0.728$&$0.045(1)$&0.108(2)&-0.015(4)\\
   16& 33.5/36&2.198\,793(4)&$1/22$&0.5&$0.728$&$0.046(1)$&0.116(5)&-0.028(9)\\
   20& 25.6/28&2.198\,795(5)&$1/22$&0.5&$0.728$&$0.046(1)$&0.12(1)&-0.04(2)\\
   24& 22.6/24&2.198\,795(7)&$1/22$&0.5&$0.728$&$0.045(1)$&0.12(2)&-0.03(3)\\
   16& 8.0/8&&&0.5&$0.72(1)$&&0.120(3)&\\
   20& 2.6/6&&&0.5&$0.71(1)$&&0.121(4)&\\
   24& 2.5/5&&&0.5&$0.71(1)$&&0.122(6)&\\
   32& 0.9/3&&&0.5&$0.71(1)$&&0.12(1)&\\
   12& 10.4/9&&&0.5&$0.70(1)$&&0.111(4)&0.04(3)\\
   16& 6.8/7&&&0.5&$0.70(1)$&&0.11(1)&0.06(5)\\
   20& 2.3/5&&&0.5&$0.70(2)$&&0.11(2)&0.1(1)\\
   24& 2.3/4&&&0.5&$0.70(4)$&&0.10(5)&0.1(2)\\
 \hline
 \hline
 \end{tabular}
 \end{center}
 \end{table*}

\begin{table*}
 \begin{center}
\caption{Fits of the finite-size Monte Carlo data of the magnetic susceptibility $\chi_0$ to (\ref{fit2}) for the critical Ising ($n=1$), XY ($n=2$) and Heisenberg ($n=3$) models.}
 \label{fit_result_2}
 \begin{tabular}[t]{llllllll}
 \hline
  \hline
   $n$ \qquad\qquad& $L_{\rm min}$ \qquad\qquad   & $\chi^2/$DF \qquad \qquad &$2y_h-d$  \qquad \qquad \qquad & ${\hat p}$  \qquad \qquad \qquad   & $q_1$  \qquad \qquad\qquad    & $q_2$ \qquad  \qquad \\
 \hline
  {\multirow{13}{*}{1}}
  &  8&6.3/9&$2$&$0.56(4)$&$1.2(1)$&0.7(2)\\
  & 10&5.0/8&$2$&$0.50(7)$&$1.4(3)$&0.5(3)\\
  & 12&1.3/7&$2$&$0.6(1)$&$1.0(3)$&1.0(3)\\
  &  8&8.2/10&$2$&$0.5$&$1.440(6)$&0.46(1)\\
  & 10&5.0/9&$2$&$0.5$&$1.448(8)$&0.45(1)\\
  & 12&3.4/8&$2$&$0.5$&$1.44(1)$&0.46(2)\\
  & 16&2.7/7&$2$&$0.5$&$1.45(1)$&0.45(3)\\
  & 20&0.8/6&$2$&$0.5$&$1.47(2)$&0.40(4)\\
  & 8&6.2/9&$2.006(5)$&$0.5$&$1.35(6)$&0.56(7)\\
  & 10&5.0/8&$2.001(7)$&$0.5$&$1.4(1)$&0.5(1)\\
  & 12&1.3/7&$2.01(1)$&$0.5$&$1.2(1)$&0.7(2)\\
  & 16&1.1/6&$2.02(1)$&$0.5$&$1.2(2)$&0.8(3)\\
  & 20&0.6/5&$2.01(2)$&$0.5$&$1.3(3)$&0.6(4)\\
 \hline
  {\multirow{8}{*}{2}}
  & 8&0.6/4&$2$&$0.45(4)$&$1.5(2)$&0.3(2)\\
  & 16&0.2/3&$2$&$0.4(1)$&$2(1)$&0(1)\\
  &  8&1.8/5&$2$&$0.5$&$1.254(6)$&0.49(1)\\
  & 16&1.1/4&$2$&$0.5$&$1.25(1)$&0.50(2)\\
  & 32&0.4/3&$2$&$0.5$&$1.23(2)$&0.54(5)\\
  & 48&0.1/2&$2$&$0.5$&$1.21(4)$&0.58(9)\\
  &  8&0.5/4&$1.995(4)$&$0.5$&$1.32(6)$&0.41(7)\\
  & 16&0.2/3&$1.99(1)$&$0.5$&$1.4(2)$&0.3(2)\\
 \hline
  {\multirow{9}{*}{3}}
  &10&26.1/11&$2$&$0.5$&$1.109(3)$&0.565(5)\\
  &12&13.4/10&$2$&$0.5$&$1.100(4)$&0.581(7)\\
  &16&8.1/8&$2$&$0.5$&$1.089(6)$&0.60(1)\\
  &20&4.0/6&$2$&$0.5$&$1.075(9)$&0.63(2)\\
  &24&4.0/5&$2$&$0.5$&$1.07(1)$&0.63(3)\\
  &28&0.5/4&$2$&$0.5$&$1.05(2)$&0.68(4)\\
  &32&0.4/3&$2$&$0.5$&$1.06(3)$&0.66(5)\\
  &16&3.3/7&$1.98(1)$&$0.5$&$1.4(1)$&0.2(2)\\
  &20&2.9/5&$1.98(2)$&$0.5$&$1.3(2)$&0.3(3)\\
 \hline
 \hline
 \end{tabular}
 \end{center}
 \end{table*}

\begin{table*}
 \begin{center}
\caption{Fits of the finite-size Monte Carlo data of the two-point correlation $g(L/2,L)$ to (\ref{fit3}) for the critical XY model.}
 \label{fit_result_3}
 \begin{tabular}[t]{lllllll}
 \hline
  \hline
     &$L_{\rm min}$ \qquad \qquad   & $\chi^2/$DF \qquad \qquad& $2y_h-2d$  \qquad \qquad \qquad & ${\hat p}$  \qquad \qquad \qquad& $v_1$  \qquad \qquad\qquad   & $v_2$ \qquad  \qquad \\
 \hline
  &24&4.0/5&$-2$&$0.5$&$0.607(3)$&0.339(6)\\
  &32&2.2/4&$-2$&$0.5$&$0.613(6)$&0.33(1)\\
  &40&1.5/3&$-2$&$0.5$&$0.61(1)$&0.34(2)\\
  &24&3.3/4&$-1.99(1)$&$0.5$&$0.54(8)$&0.42(9)\\
  &32&1.3/3&$-2.02(2)$&$0.5$&$0.8(2)$&0.1(3)\\
  &40&1.3/2&$-2.02(4)$&$0.5$&$0.8(4)$&0.1(6)\\
  &16&28.5/5&$-2$&$0.52(5)$&$0.55(8)$&0.41(9)\\
  &24&4.8/4&$-2$&$0.5(1)$&$0.7(3)$&0.2(3)\\
 \hline
 \hline
 \end{tabular}
 \end{center}
 \end{table*}

\bibliography{papers}